\documentclass[longbibliography,aps,prb,twocolumn,superscriptaddress]{revtex4-2}

\usepackage{bm}
\usepackage{amssymb}
\usepackage{amsmath}
\usepackage{siunitx}
\usepackage{textcomp}
\usepackage{graphicx}
\usepackage{natbib}
\usepackage{color}
\usepackage{setspace}
\usepackage[dvipdfmx]{hyperref}
\hypersetup{colorlinks=true, citecolor=blue, urlcolor=blue, linkcolor=blue}
\graphicspath{{./figures/}}

\begin{document}

\title{Dynamics of pinned quantized vortices in superfluid $^4$He in a microelectromechanical oscillator}


\author{Tomo Nakagawa}
\affiliation{Department of Physics, Osaka City University, 3-3-138 Sugimoto, 558-8585 Osaka, Japan}
\author{Makoto Tsubota}
\affiliation{Department of Physics, Nambu Yoichiro Institute of Theoretical and Experimental Physics(NITEP), Osaka Metropolitan University, 3-3-138 Sugimoto, 558-8585 Osaka, Japan}
\author{Keegan Gunther}
\affiliation{Department of Physics, University of Florida, Gainesville, FL 32611, USA}
\author{Yoonseok Lee}
\affiliation{Department of Physics, University of Florida, Gainesville, FL 32611, USA}


\date{\today}

\begin{abstract}
We numerically studied the vortex dynamics at zero temperature in superfluid $^4$He confined between two parallel rough solid boundaries, one of which oscillates in a shear mode. This study was motivated by the experimental work by Barquist $et$ $al.$ which employed a microelectromechanical systems (MEMS) oscillator operating in superfluid $^4$He at a near-zero temperature. Their experiments suggest that the motion of the MEMS oscillator is damped by quantized vortices. In our study, we postulated that this damping effect was closely associated with vortex pinning phenomena and developed pinning models. 
Our primary objective is to understand the vortex dynamics in the presence of pinning and to provide insight into the experimental observations regarding the damping mechanism.
We confirmed that Kelvin waves were excited in the pinned vortices when the oscillation frequency of the solid boundary matched with the mode frequency of the Kelvin wave. Additionally, we examined the formation and evolution of vortex tangles between the boundaries. The vortex tangle was suppressed in the presence of pinning, while the absence of pinning allowed to form well developed vortex tangle resulting in turbulence. Finally, by evaluating the tension of pinned vortices we extracted the damping force acting on the solid boundaries.
\end{abstract}


\maketitle

\section{introduction}\label{sec:introduction}
Vortex pinning is a prevalent phenomenon observed in various quantum condensed matter systems, including superfluid helium, superconductors \cite{Tilley,Tinkham,Gedik1995,Friesen1996,Matsushita,Shklovskij,Nagai2022}, and neutron stars \cite{Anderson1975,Alpar1981,Alpar1993,Alpar1996}.  Pinning of quantized vortices can occur due to inhomogeneities present on the boundaries or in the volume. This pinning significantly influences the motion of the vortices, alongside the dynamics within the bulk system, sometimes resulting in practical consequences such as the critical current in superconducting wires.   In superfluid $^4$He, the phenomenon of quantized vortex pinning arises from the presence of surface roughness, notably in the form of bumps on solid boundaries \cite{Glaberson1966,Schwarz1981,Adams1985,Schwarz1985,Schwarz1992,Schwarz1993,Tsubota1993,Tsubota1994,Hakonen1998,Zieve2012,Neumann2014,Stagg2017}.  While a significant body of research has been dedicated to investigating the dynamics of quantized vortices in a bulk system, such as quantum turbulence generated by thermal counterflow \cite{Vinen19571,Vinen19572,Vinen19573,Vinen19574,Schwarz1988,Adachi2010}, it is imperative to recognize that local effects resulting from surface roughness and pinning cannot be disregarded in some cases. The core of a quantized vortex in superfluid $^4$He is so thin ($\sim 0.1\ \mathrm{nm}$) that any realistic surface is rough for quantized vortices, and consequently, pinning is ubiquitous. Experimental investigations have confirmed the influence of pinning and surface roughness in various phenomena, including phase slip phenomena \cite{Glaberson1966,Hakonen1998}, spin-up \cite{Adams1985}, and vortex motion along the surfaces \cite{Zieve2012}. In numerical studies, the pinning and surface roughness have been studied using the vortex filament model \cite{Schwarz1981,Schwarz1985,Schwarz1992,Schwarz1993,Tsubota1993,Tsubota1994} and the Gross-Pitaevskii model \cite{Stagg2017}.

Recently, the experimental group at the University of Florida (Florida group) observed interesting phenomena that, they claim, are related to pinning \cite{Barquist2020,Barquist2022}. 
They employed microelectromechanical systems (MEMS) for experiments in both $^3$He \cite{Gonzalez2016,Zheng2017,Zheng20172}  and $^4$He \cite{Gonzalez2013,Gonzalez20132,Barquist2020,Barquist2022,lancaster,aalto}.
MEMS are mechanical devices with dimensions in the micro-meter scale and low mass, affording them high sensitivity for force, position, and mass sensing \cite{Bachtold2022}. 
The MEMS device of the Florida group consists of a  $125\times 125$~$\mu$m$^{2}$ thin plate, suspended above a substrate by serpentine springs forming a uniform gap of 2~$\mu$m between them. \cite{Barquist2020}.  This device is a mechanical oscillator in which the thin plate oscillates in a shear mode at its resonance frequency of $\approx 24$~kHz.
The Florida group investigated the influence of quantized vortices on the damping of the MEMS by measuring its response in superfluid $^4$He at 14~mK, both with and without quantum turbulence being actively generated by a nearby quartz tuning fork (QTF). They observed a few interesting and new phenomena. The first was the unexpectedly higher damping than what was expected at this low temperature. The second was a phenomenon that the Florida group called annealing: a hysteresis in damping as a function of device velocity, {\it i.e.} oscillation amplitude. The annealing effect was only observed in the absence of quantum turbulence, and disappeared when turbulence was actively generated. The third was the enhanced phase noise in the presence of quantum turbulence.

They argue that the underlying cause of all of these phenomena is associated with vortex pinning and the vortex-vortex interaction. If the vortices, trapped between the plate and substrate, are not pinned, they just slip along the plate's surface and do not impede its motion. However, in the presence of pinning, the vortices are stretched by the plate's motion, resulting in the transfer of oscillation energy to the vortices and leading to damping. They also claimed that this energy transfer was efficient when the two mechanical systems -- the oscillator and the vortex line -- are resonantly coupled.

When measuring the hysteresis, the Florida group prepared a quiescent state of superfluid helium, in which no turbulence was actively generated, but remnant vortices were present \cite{Awschalom1984}. Then the velocity of the oscillator was gradually increased, and then decreased back to its original value, while measuring the damping in the oscillator. It was found that the damping decreased after the excitation sweep, and remained lower until quantum turbulence was generated near the device.

The Florida group provided an explanation for this hysteresis  in terms of the remnant vortices. During the upward sweep, the remnant vortices that are trapped and pinned in the gap contribute to the damping. As the upward sweep progresses, some of these remnant vortices are eliminated, resulting in a reduction in the damping during the downward sweep. This elimination is thought to be promoted by a combined processes of vortex depinning and annihilation of vortex anti-vortex pairs. This was further supported by repeating the experiment in the presence of quantum turbulence generated by a QTF. With the continuous injection of additional vortex rings, the system reached a statistically steady state and erased the hysteresis.

In this paper, we develop a couple of models for vortex pinning to understand the vortex dynamics and the experimentally observed phenomena. We attempt to describe the vortex dynamics in such a system and perform single and multiple vortex simulations to elucidate the effects of vortex interaction. 


\section{formulation}\label{sec:formulation}
Dynamics of quantized vortices is described by the vortex filament model \cite{Schwarz1981,Schwarz1985} where a vortex line is modeled as a 1-dimensional filament. This model cannot describe phenomena related to its finite core such as vortex creation/annihilation and vortex reconnection.   However, it has been highly successful in revealing fundamental physics of vortex dynamics in superfluid $^4$He \cite{Tsubota2013,Tsubota2017}.

The motion of vortex filaments at $\bm{s}(\xi_0)$ obeys the Biot-Savart law: 
\begin{equation}\label{eq:BS}
	\frac{d\bm{s}(\xi_0)}{dt}=\frac{\kappa}{4 \pi} \int_{\mathcal{L}} \frac{\mathbf{s}^{\prime}(\xi) \times\left[(\mathbf{s}(\xi)-\mathbf{s}(\xi_0)\right]}{|\mathbf{s}(\xi)-\mathbf{s}(\xi_0)|^{3}} d \xi+\bm{v}_{s,b},
\end{equation}
where $\kappa$ and $\bm{s}(\xi)$ denote the quantum circulation and the position of the vortex filaments parameterized by $\xi$; $\bm{s}^\prime$ refers to $d\bm{s}/d\xi$; $\bm{v}_{s,b}$ is the velocity induced by boundaries. The Biot-Savart integration is performed over all vortex filaments ${\mathcal{L}}$.

To numerically calculate the equation of motion, a vortex filament is discretized as connection of points along the filament. The separation of the adjacent points,  $\Delta \xi = s_{i+1} - s_{i}$ is set between $\Delta \xi_\mathrm{min}$ and  $\Delta \xi_\mathrm{max}$. The equation of motion is solved by the fourth-order Runge-Kutta scheme. Because the reconnection events cannot be described directly in this model, the vortices are algorithmically reconnected when the distance of vortices becomes less than $\Delta \xi_\mathrm{min}$.  Short vortices consisting of less than 6 points are removed from the system resulting in dissipation.  This is the main channel for dissipation in the simulation.  The proposed dissipation mechanism is justified because such short vortex lines are isolated from the other vortices. Furthermore, such short vortex lines can never grow to influence the vortex dynamics since their Kelvin wave modes are at much higher frequencies compared to the oscillation frequency. This aspect will be discussed further later.

In addition to the inherent dynamics within the bulk, it is imperative to account for the intricate effects of pinning. We employ two models to incorporate pinning in the simulation: the critical angle model (Sec. \ref{sec:cam}) and the hemispherical pinning site model (Sec. \ref{sec:hps}). The former model, introduced by Schwarz \cite{Schwarz1993}, successfully described the effect of pinning on vortex dynamics particularly for vortices partially attached to a wire in accordance with the findings of Zieve $et$ $al.$ \cite{Zieve1992}. It effectively captures the dynamics in the presence of highly irregular boundaries. While the critical angle model may not provide an explicit description of pinning and depinning events, it remains valuable in characterizing vortex dynamics on rough surfaces, irrespective of the intricacies of surface geometry.
 On the other hand, the latter model addresses the dynamics of vortices in the presence of hemispherical pinning sites through exact solutions of boundary conditions. Schwarz \cite{Schwarz1985} explored the static boundary conditions of (hemi)spherical surfaces using the vortex filament model. Fujiyama $et$ $al.$ \cite{Fujiyama2009} resolved the boundary conditions for a moving sphere using the vortex filament model, enabling the study of turbulence generation. This model offers insights into the pinning and depinning phenomena, although it encounters challenges when dealing with boundaries featuring multiple irregularities.
 Utilizing these two models, we analyze the dynamics of pinned vortices in the system.

\subsection{Critical angle model}\label{sec:cam}

\begin{figure}
	\includegraphics[width=\linewidth]{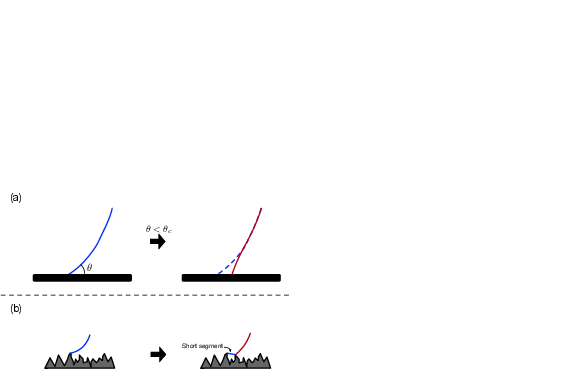}
	\caption{\label{fig:critical} Basic idea of the critical angle model. (a) When the angle between the pinned vortex and the solid boundary is less than the critical value, the end of the vortex jumps to another point and pinned again. (b) The sufficiently tilted vortex reconnects with another bump on the rough surface. The vortex splits into a short segment and a line vortex shorter than the original one. The segment is ignored in this model.}
\end{figure}

 The ends of a vortex are forcibly fixed on the boundary when pinned. Except for the ends, the motion of the vortex is governed by the equation of motion, Eq.~(\ref{eq:BS}).  When the vortex is in motion, the vortex line tilts away from its static angle, normal to the boundary.  The vortex undergoes depinning at a critical angle, $\theta_{c}$. Here, we use $\theta_c=\pi/6$. The specific choice of $\theta_{c}$ does not influence the results of the simulation \cite{critical_sup}. The depinned vortex promptly becomes pinned again by another nearby point, analogous to the scenario depicted in Fig. \ref{fig:critical} (a). The exact location of this "repinning" point is determined by a point $\bm{s}_i$ adjacent to a boundary and a neighboring point $\bm{s}_{i+1}$ on the opposite side of the boundary: the intersection point on the boundary by the line drawn perpendicularly from the midpoint between $s_{i}$ and $s_{i+1}$. In simulation, the end of the vortex appears to jump from one position to another. 

This algorithm allows a physical representation of vortex dynamics in the presence of a rough surface, Fig.~\ref{fig:critical} (b).  When a vortex pinned on one of the bumps tilts beyond a critical angle, it undergoes a reconnection process with an adjacent bump. This reconnection event results in the splitting of the vortex into two: a short segment vortex bridging the bumps and the original vortex shortened by the length of the short segment. In this model, we ignore the influence of the short segment because Kelvin waves cannot be excited in this short segment, and it eventually dissipates away through the repeated reconnection events. We refer to this mechanism as rough surface-induced dissipation (RSID).  In our simulation, we assume an extremely rough surface, and the vortex orientation is not normal to the averaged flat boundary. Although the precise velocity field in the vicinity of the boundary is not accurately described, we believe that velocity fluctuations at the scales smaller than the spatial resolution have minimal impact on the dynamics of the whole system. 
\begin{table*}
	\centering
	\begin{tabular}{l|c|l}
		Parameters in the critical angle model& Symbol & Value \\ \hline\hline
		$\bm{\mathrm{Single\ vortex\ case} }$& &  \\
		Distance between plates & $L$ & $2\ \mathrm{\mu m}$ \\
		Time resolution ($n=1$ mode)  & $\Delta t$ & $2.5\ \mathrm{ns}$\\
		Minimum spatial resolution ($n=1$ mode) & $\Delta \xi_\mathrm{min}$ & $0.01\ \mathrm{\mu m}$\\
		Maximum spatial resolution ($n=1$ mode) & $\Delta \xi_\mathrm{max}$ & $0.03\ \mathrm{\mu m}$\\
		Time resolution ($n=3$ and $n=10$ mode)  & $\Delta t$ & $0.5\ \mathrm{ns}$\\
		Minimum spatial resolution ($n=3$ and $n=10$ mode) & $\Delta \xi_\mathrm{min}$ & $0.004\ \mathrm{\mu m}$\\
		Maximum spatial resolution ($n=3$ and $n=10$ mode) & $\Delta \xi_\mathrm{max}$ & $0.01\ \mathrm{\mu m}$\\
		Oscillation frequency ($n=1$ mode) & $f_1$ & $23625\ \mathrm{Hz}$\\
		Oscillation frequency ($n=3$ mode)& $f_3$ & $23625 \times 3^2 \ \mathrm{Hz}$\\
		Oscillation frequency ($n=10$ mode)& $f_{10}$ & $23625 \times 10^2 \ \mathrm{Hz}$\\
		Oscillation amplitude & A & $0.1\ \mathrm{\mu m}$\\ \hline
		$\bm{\mathrm{Multiple\ vortex\ case} }$& &  \\
		Distance between plates & $L$ & $2\ \mathrm{\mu m}$ \\
		Time resolution   & $\Delta t$ & $2.5\ \mathrm{ns}$\\
		Minimum spatial resolution  & $\Delta \xi_\mathrm{min}$ & $0.01\ \mathrm{\mu m}$\\
		Maximum spatial resolution & $\Delta \xi_\mathrm{max}$ & $0.03\ \mathrm{\mu m}$\\
		Oscillation frequency ($n=10$ mode)& $f$ & $23625\ \mathrm{Hz}$\\
		Oscillation amplitude & A & $0.1\ \mathrm{\mu m}$\\ \hline
	\end{tabular}
	\caption{Parameters in the critical angle model }
	\label{tb:result1table}
\end{table*}

\subsection{Hemispherical pinning site model}\label{sec:hps}
In this model, hemispherical pinning sites are added on a flat boundary. This model can describe the depinning events and the dynamics of the short segment.  The superfluid velocity normal to the pinning site should vanish. The boundary condition of a static sphere is solved in terms of the associated Legendre polynomial \cite{Schwarz1985}.  Moreover, when the pinning site moves, a potential flow $\bm{v}_u(\bm{r})$ is applied \cite{Fujiyama2009}:
\begin{align}
	\Phi_{u}(\boldsymbol{r})&=-\frac{1}{2}\left(\frac{a}{r}\right)^{3} \bm{u}_{p} \cdot \boldsymbol{r},\\
	\boldsymbol{v}_{u}(\boldsymbol{r})&=\nabla \Phi_{u}(\boldsymbol{r}).
\end{align}
Here, $a$ and $\bm{u}_p$ are the radius and velocity of the pinning site. We use $a = 0.03\ \mathrm{\mu m}$ \cite{size}.

The vortices that come close enough to flat boundaries or pinning sites, specifically within $\Delta \xi_\mathrm{min}$, undergo reconnection with them. When a part of a pinned vortex gets sufficiently close to the flat boundary, it reconnects with the boundary and becomes free from pinning; this phenomenon is referred to as depinning in our model. The reverse is also true; a free vortex can reconnect with a pinning site and gets pinned on it.
The model itself does not yield dissipation. However, the dissipation mechanism on the rough surface, such as RSID, can be incorporated in this model without algorithmically removing the contribution of short vortex segments.

\begin{table*}
	\centering
	\begin{tabular}{l|c|l}
		Parameters in the hemispherical pinning site model& Symbol & Value \\ \hline\hline
		Distance between plates & $L$ & $2\ \mathrm{\mu m}$ \\
		Pinning site size & $a$ & $0.03\ \mathrm{\mu m}$ \\
		Time resolution& $\Delta t$ & $0.05\ \mathrm{ns}$\\
		Minimum spatial resolution& $\Delta \xi_\mathrm{min}$ & $0.001\ \mathrm{\mu m}$\\
		Maximum spatial resolution& $\Delta \xi_\mathrm{man}$ & $0.003\ \mathrm{\mu m}$\\
		Oscillation frequency ($n=1$ mode) & $f$ & $23625\ \mathrm{Hz}$\\
		Oscillation amplitude & A & $0.1\ \mathrm{\mu m}$\\ \hline
	\end{tabular}
	\caption{Parameters in the hemispherical pinning site model.}
	\label{tb:result2table}
\end{table*}

\subsection{System}

We study numerically the experimental configuration using the pinning models described above. Specifically, we examine the behavior of pinned vortices situated between two parallel solid boundaries realized in the MEMS oscillator. The separation distance, $L$, between the boundaries is $2\ \mathrm{\mu m}$, and one of the boundaries oscillates in the $x$-direction with the velocity vector $\bm{v}(t)=2\pi f A \cos(2\pi f t)\hat{\bm{x}}$. Here, $A$ represents the amplitude of the oscillation, while $f$ denotes the oscillation frequency. The pinned ends of the vortices also oscillate in the same manner.  The excitation of the vortex line is described by the Kelvin waves whose dispersion relation is given by
\begin{align}\label{eq:resonant}
	f(k)=\frac{1}{2\pi}\frac{\kappa k^2}{4\pi}\ln\left(\frac{1}{ka_0}\right),
\end{align}
where $a_0$ and $k$ are a vortex core size and the wavenumber of the Kelvin wave \cite{Donnelly}. In our simulation, unless otherwise mentioned, we use $f_{1} = 23625$~Hz for the oscillation frequency of the moving boundary.  This frequency is the first resonant frequency of the Kevin wave mode  specified by $k_{n} = n\pi /L$ with $n = 1$, which is close to the value used in the experiment \cite{Barquist2020}.   

\section{Results}
We present the results of the simulations in two cases: single vortex case (Sec.~\ref{sec:single}) and multiple vortex case (Sec.~\ref{sec:multi}). Subsequently, in Section \ref{sec:comp}, we revisit the experimental observations in an effort to make physical connections with our simulations. The specific parameters used in the simulations are outlined in Table \ref{tb:result1table} and Table \ref{tb:result2table}.

\subsection{Vortex dynamics with pinning}\label{sec:Vortex_dynamics}

\begin{figure}
	\includegraphics[width=\linewidth]{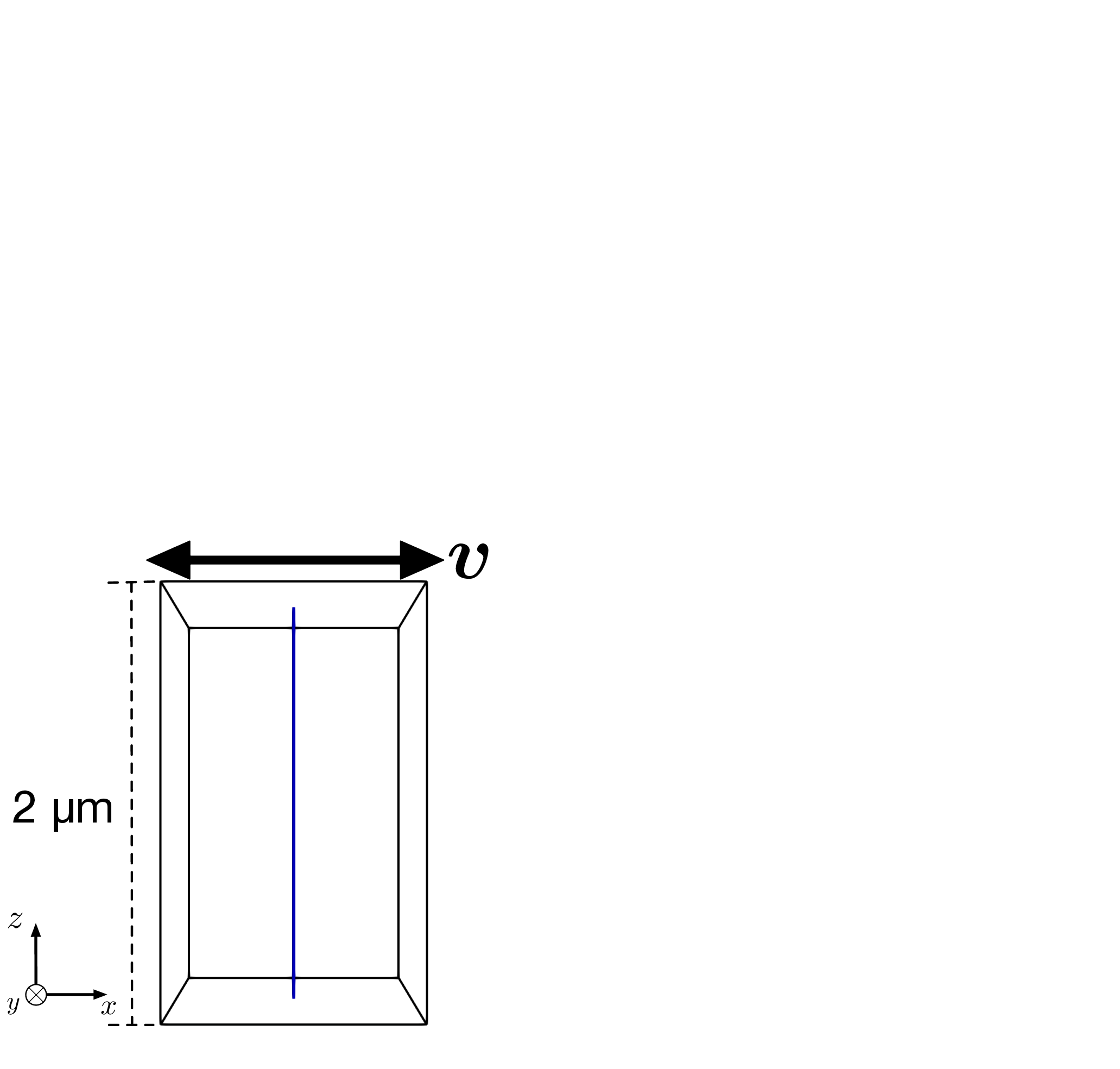}
	\caption{\label{fig:initial_single}Initial state of a single vortex trapped between two parallel solid boundaries.  The arrow indicates the direction of oscillation of the top boundary.}
\end{figure}

\subsubsection{Single vortex case}\label{sec:single}

\begin{figure}
	\includegraphics[width=\linewidth]{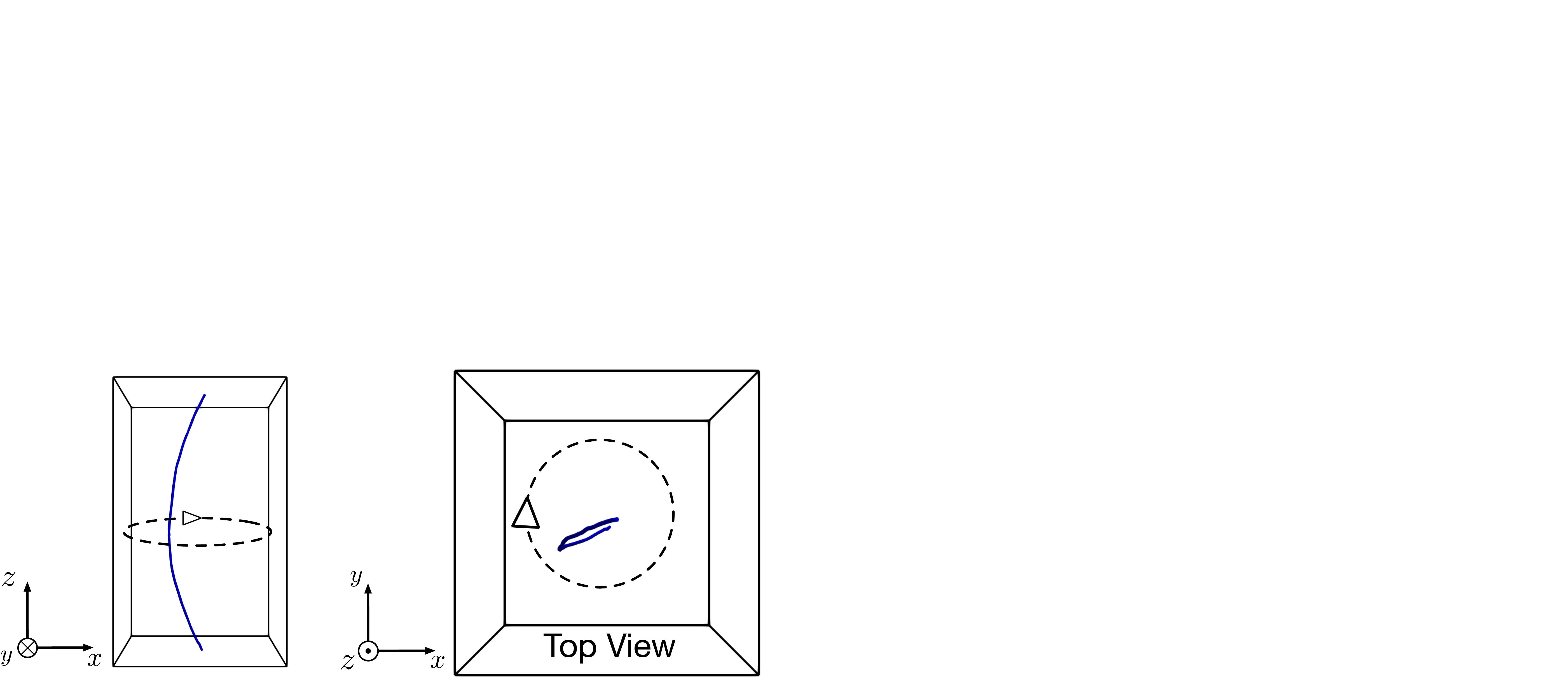}
	\caption{\label{fig:singlesteady}Snapshots of the statistically steady state in the single vortex case. The first Kelvin wave mode is excited, and the vortex continues to rotate by its self-induce velocity. The arrows indicate the rotation direction. }
\end{figure}

\begin{figure}
	\includegraphics[width=\linewidth]{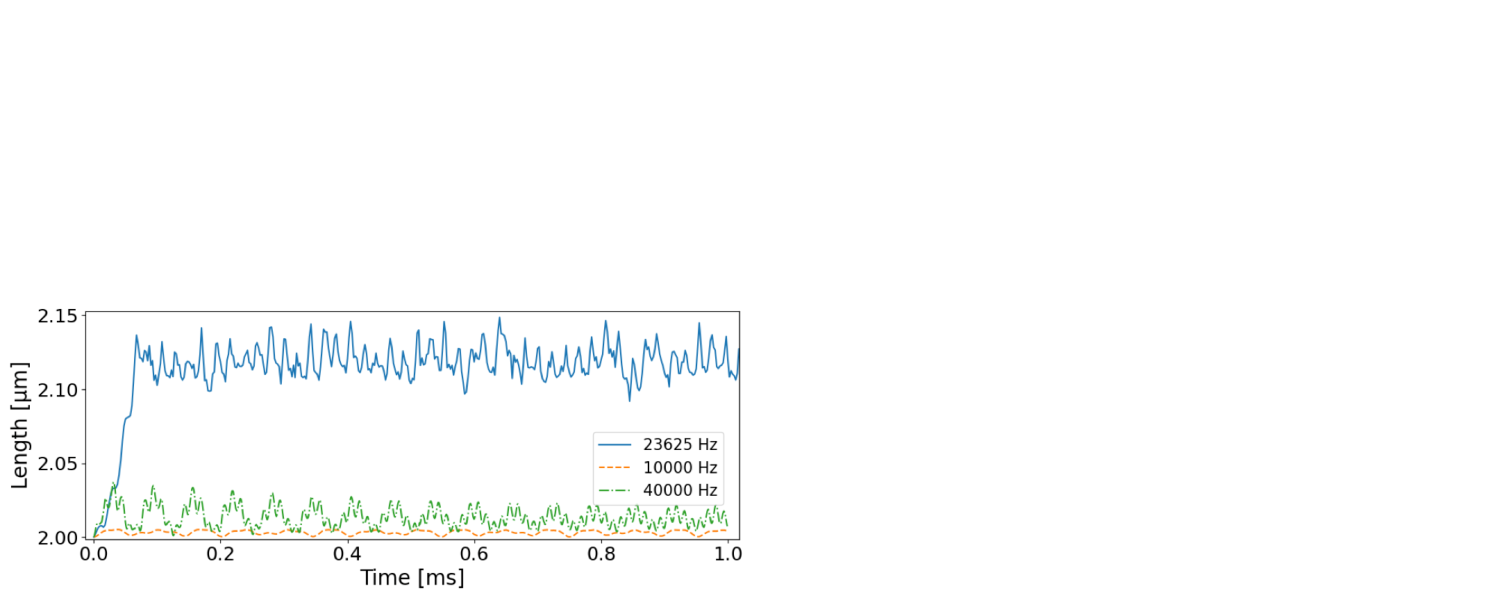}
	\caption{\label{fig:singlelength}Time development of the vortex line length in the single vortex case for three different oscillation frequencies. If the excitation frequency does not match with the Kelvin wave resonance, the length of the vortex does not increase. }
\end{figure}

First, we perform single vortex simulations. In this case,  a rectilinear vortex is placed between the solid boundaries as an initial state as shown in Fig. \ref{fig:initial_single}. Both ends of the vortex are pinned on the boundaries. The orientation of the vortex is from bottom to top along the $z$-direction, $+\hat{\bm{z}}$.
Assuming the presence of rough boundaries on both sides, we apply the critical angle model to describe their behavior. As the upper boundary initiates oscillation, the vortex becomes stretched and curved gradually. Simultaneously, the vortex begins to rotate around its initial configuration due to its self-induced velocity. Looking from the top, the rotation direction is clockwise reflecting the vortex orientation as depicted in Fig. \ref{fig:singlesteady}. The temporal evolution of the vortex line length is illustrated in Fig. \ref{fig:singlelength} for the amplitude of $A = 0.1\ \mathrm{\mu m}$. When the frequency $f$ matches the resonant value of $23625\ \mathrm{Hz}$, the vortex line length increases from its initial value of $2\ \mathrm{\mu m}$ until reaching a statistically steady state of approximately $2.13\ \mathrm{\mu m}$. This steady state corresponds to the Kelvin wave of the $n=1$ mode. The vortex continues to rotate due to its self-induced velocity with the oscillation period of $1/f$.

\begin{figure}
	\includegraphics[width=\linewidth]{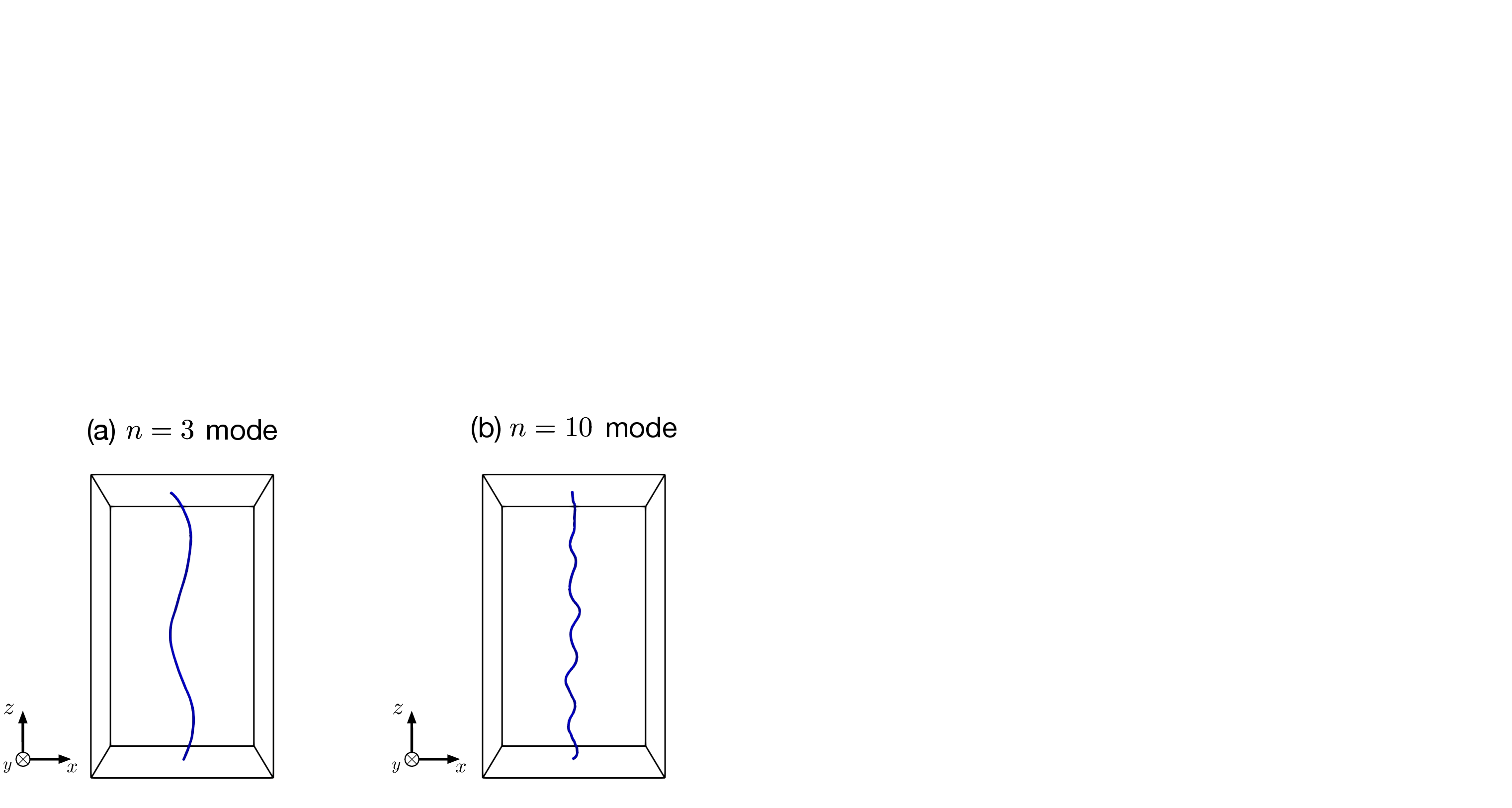}
	\caption{\label{fig:mode}Snapshots of the single vortex when oscillated at the frequencies of the (a) $n=3$ mode and (b) $n=10$ mode. Each clearly shows the corresponding Kelvin wave excited in the vortex. Both snapshots were taken at $t = 0.027$~ms.}
\end{figure}

The statistically steady state represents an equilibrium between excitation and dissipation mechanisms. The excitation arises from the driven oscillation, while the dissipation comes from the rough surface-induced dissipation (RSID) discussed in Sec. \ref{sec:cam}. As the vortex is sufficiently tilted, it reconnects with another bump, effectively shortening in length. Repeating this process, the vortex gets shorter. When the excitation from the oscillation is counterbalanced by RSID, the system reaches a statistically steady state. The simulation clearly confirms the effectiveness of RSID.

Next, we examine the frequency dependence of vortex dynamics. When the system is driven at the resonant frequency $f$ given by Eq.~(\ref{eq:resonant}), it reaches a steady state corresponding to the mode. Figure \ref{fig:mode} shows the snapshots of the steady states for the $n=3$ and $n=10$ modes. On the other hand, when the oscillation frequency does not match with a resonant frequency, the degree of stretching in the vortex diminishes, as observed in Fig.~\ref{fig:singlelength}.

\begin{figure}
	\includegraphics[width=\linewidth]{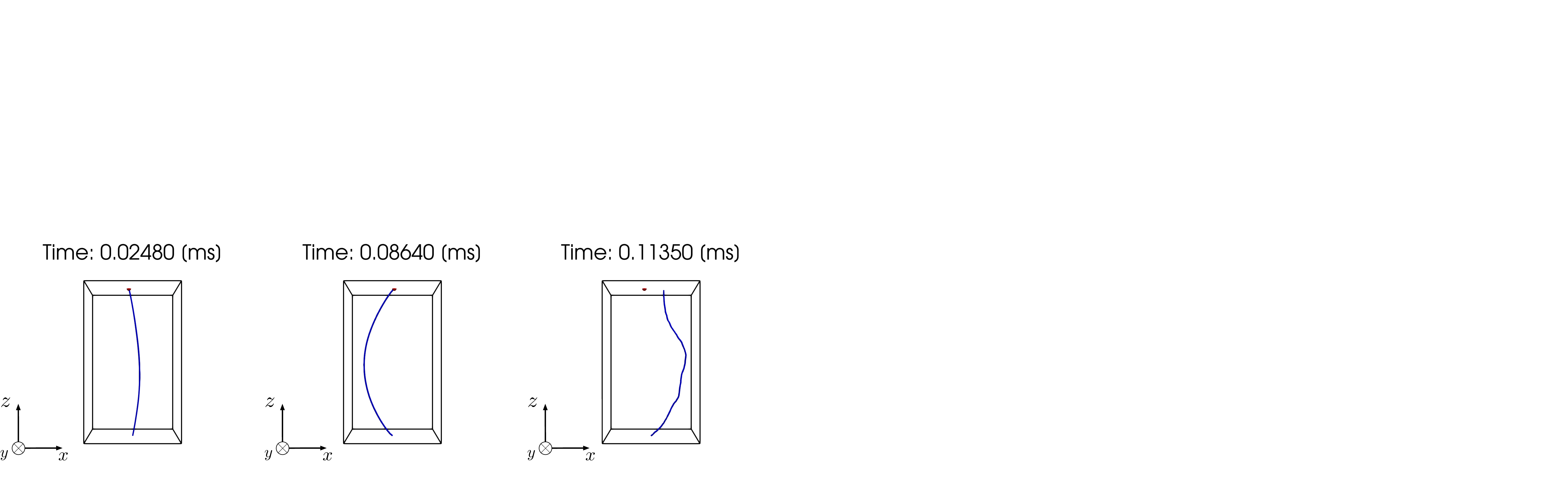}
	\caption{\label{fig:singlesite}Snapshots of the single vortex case in the  hemispherical pinning site model. The red spots on the upper and bottom boundaries refer to the pinning sites. The vortex is gradually stretched and, eventually depinned from the upper pinning site.}
\end{figure}

In contrast, the simulation in the hemispherical pinning site model demonstrates the depinning process.  With only one pinning site on each boundary, the vortex is also gradually stretched, exciting the first Kelvin wave mode.  However, the vortex does not reach a steady state as it lacks dissipation. Consequently, the vortex continues to elongate, causing the angle between the boundary and the vortex to decrease. Eventually, the vortex reconnects with the boundary and becomes free from pinning. Figure~\ref{fig:singlesite} shows the snapshots at the three different stages discussed. 

\begin{figure}
	\includegraphics[width=\linewidth]{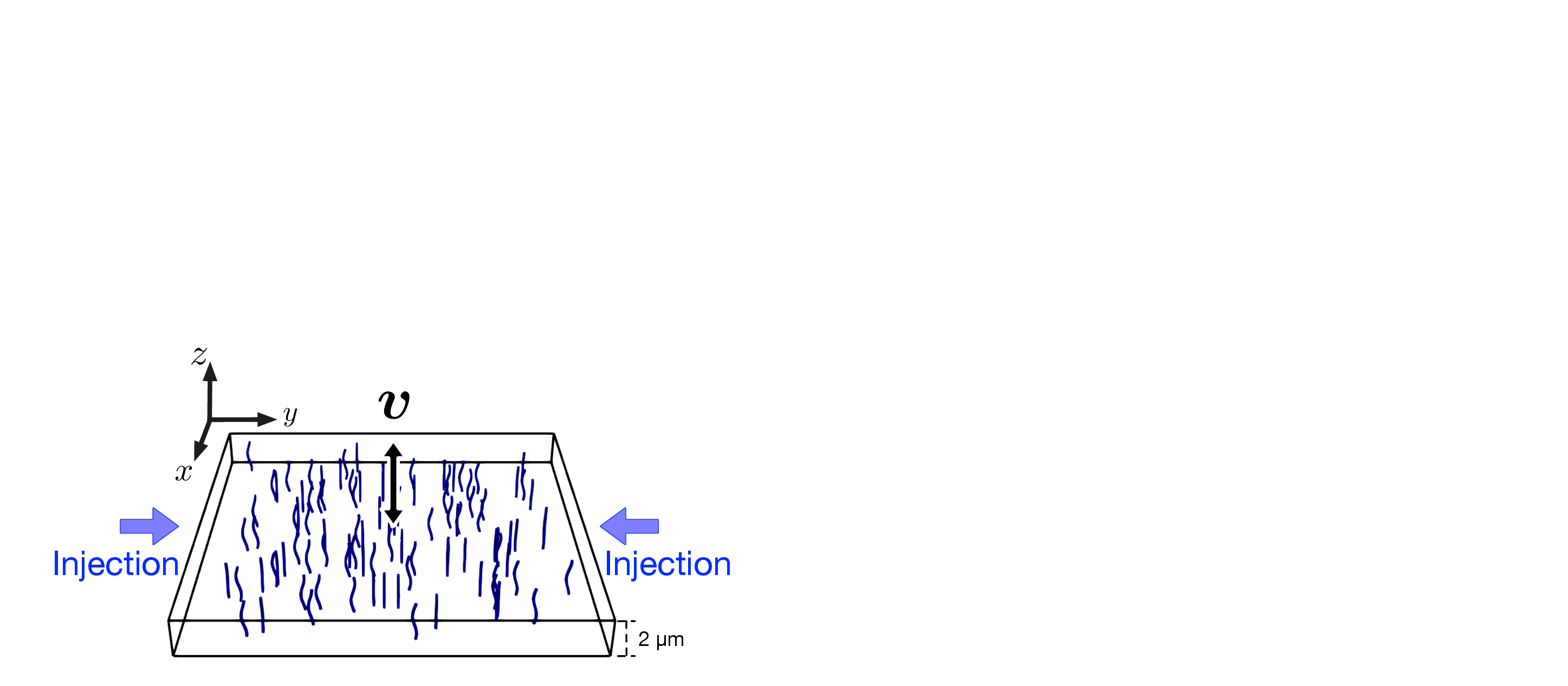}
	\caption{\label{fig:initial_multiple}Initial state of the multiple vortex simulation in the critical angle model. 100 vortices are prepared randomly with the equal probability of up or down orientation. The upper boundary oscillates in the $x$ direction. The vortices are injected from the both sides along the $y$ direction. See the text.}
\end{figure}

\subsubsection{Multiple vortex case}\label{sec:multi}

Since multiple vortices must be present in the real system, we extend our simulation involving 100 line vortices within a volume $V = 20\times 20\ \mathrm\times 2\ \mathrm{{\mu m}^3}$ defined by the two boundaries. The orientations of the vortices are initially prepared randomly with the half of them oriented upward ($+\hat{\bm{z}}$) and the other half oriented downward ($-\hat{\bm{z}}$). The ends of the vortices are pinned on the surfaces as described by the critical angle model. The lower boundary remains stationary, while the top boundary is in motion. The vortices located outside the $2\ \mathrm{\mu m}$ boundary from the volume $V$ are removed as they hardly affect the vortex dynamics of interest. 

To simulate the experiment \cite{Barquist2020,Barquist2022} where the vortex rings were produced by a nearby QTF, we inject vortices into the gap between the boundaries from the opposite sides in the $y$-direction. The blue arrows in Fig.~\ref{fig:initial_multiple} indicate the direction of the ring injection. The position of entrance into the volume is randomly determined. The vortex ring injection is characterized by two parameters: the injection frequency and the size distribution.  The injection frequency is set to $10^6\ \mathrm{Hz}$, significantly higher than the oscillation frequency of the QTF, which is the order of $10^4\ \mathrm{Hz}$.  
The high injection frequency is justified by our previous simulation in which we investigated the emission of vortices from a localized vortex tangle generated by the injection of vortex rings \cite{Nakagawa2020}. In that simulation, the local vortex tangle was formed through the collisions of the vortex rings injected into the system. We verified that the frequency of vortex ring emission from the local tangle exceeded the actual injection frequency.

\begin{figure}
	\includegraphics[width=\linewidth]{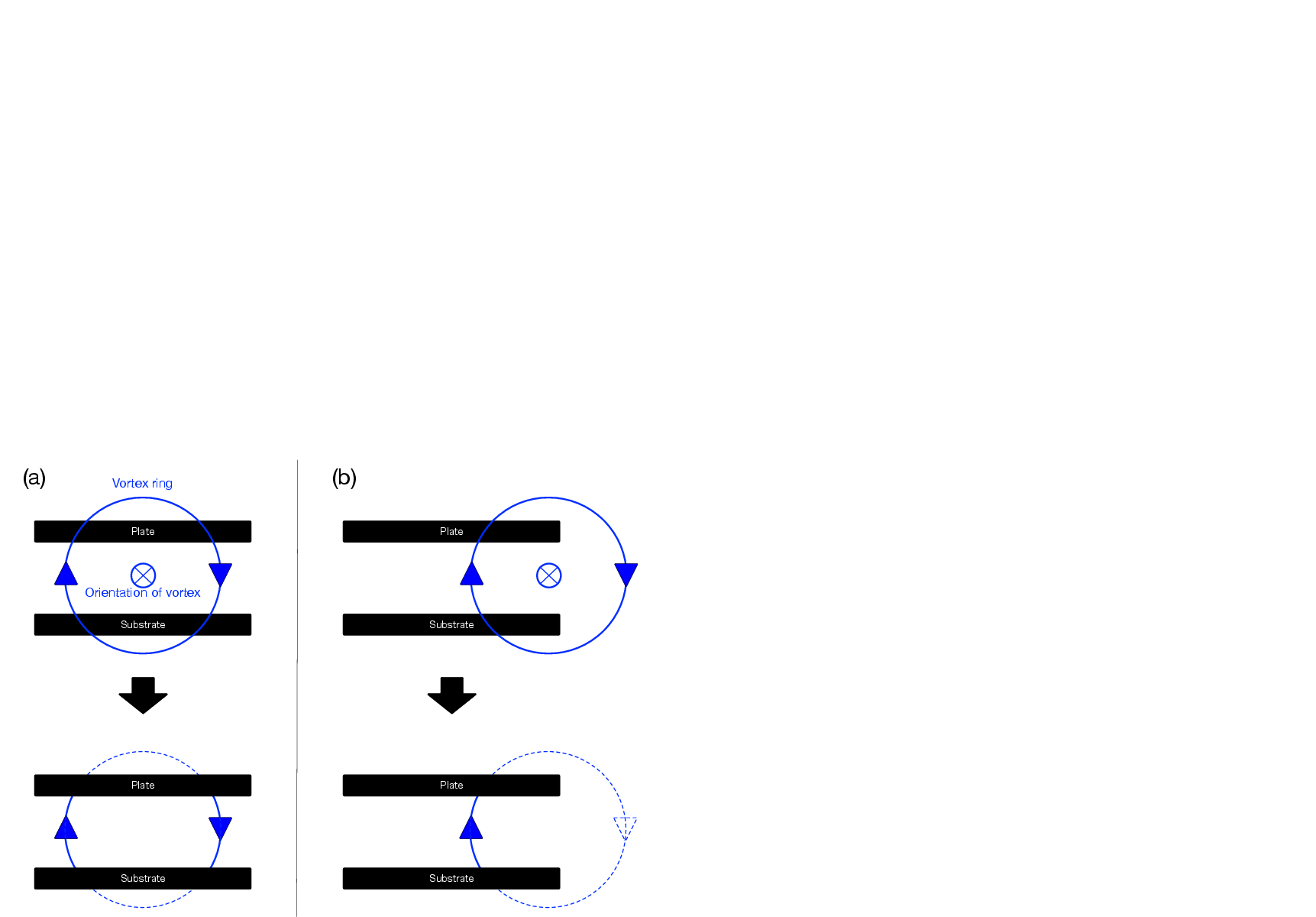}
	\caption{\label{fig:platesub}Schematic diagrams for vortex line injection in the simulation shown in Fig.~\ref{fig:initial_multiple}. When a  large vortex is injected, the vortex may interact with the boundaries to produce line vortices bridging between them. In (a), the vortex undergoes reconnection at two distinct points on the plate and the substrate, resulting in the formation of an antiparallel line vortex pair.  In (b), the vortex undergoes reconnection at only one point on each boundary, leading to the formation of a single line vortex within the gap.}
\end{figure}

\begin{figure}
	\includegraphics[width=\linewidth]{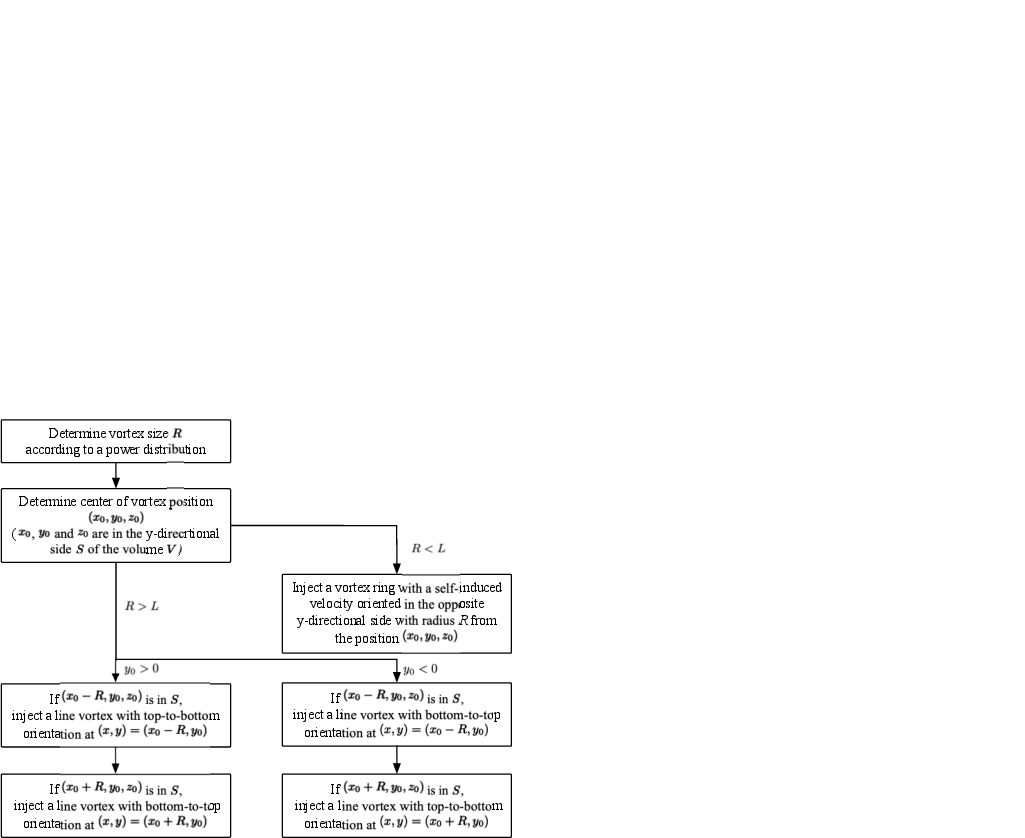}
	\caption{\label{fig:flowring}Flowchart for determining the position and shape of the injected vortices in the simulation of Fig. \ref{fig:initial_multiple}. The parts after $R>L$ corresponds to the large vortex injections illustrated in Fig. \ref{fig:platesub}.}
\end{figure}

Additionally, we assume that the size distribution of the emitted vortices from a vortex tangle follows a power-law. This power law was confirmed in both the experiment \cite{Yano2019} and the numerical work \cite{Nakagawa2020}. Yano $et$ $al.$ experimentally extracted the power-law statistics of the size of vortices emitted from a vortex tangle generated by a vibrating wire \cite{Yano2019}, which was also confirmed numerically \cite{Nakagawa2020}. In this case, the radius of the vortices is ranging from $0.25\ \mathrm{\mu m}$ to $10\ \mathrm{\mu m}$, following the power-law distribution with the exponent -1.7 observed in Ref. \cite{Yano2019}.  However, since this distribution is cumulative, we applied an exponent of $-2.7$ obtained by differentiating the distribution by the size.  Although the shape and oscillation frequency are expected to depend on  the tuning forks or the generators used, we apply the same power distribution \cite{Yano2019} because this is the only study on the size distribution of the vortex loops emitted from a local vortex tangle made by a vibrating object. The vortex rings entering the volume interacts with the existing vortices and the boundaries in two different ways.  If the diameter of a ring is smaller than the plate gap of $2\ \mathrm{\mu m}$, the vortex ring enters into the system freely and interact with the existing vortices. However, if the diameter exceeds the gap, the ring interacts with the boundaries.  In this case the injected vortex may be converted into a line vortex pair (Fig.~\ref{fig:platesub}(a)) or a single line vortex (Fig.~\ref{fig:platesub}(b)) inside the gap. Thus, we effectively inject a pair of antiparallel line vortices (Fig. \ref{fig:platesub}(a)) or a single line vortex (Fig. \ref{fig:platesub}(b)), respectively. The flowchart for the vortex injection is shown in Fig. \ref{fig:flowring}.

\begin{figure}
	\includegraphics[width=\linewidth]{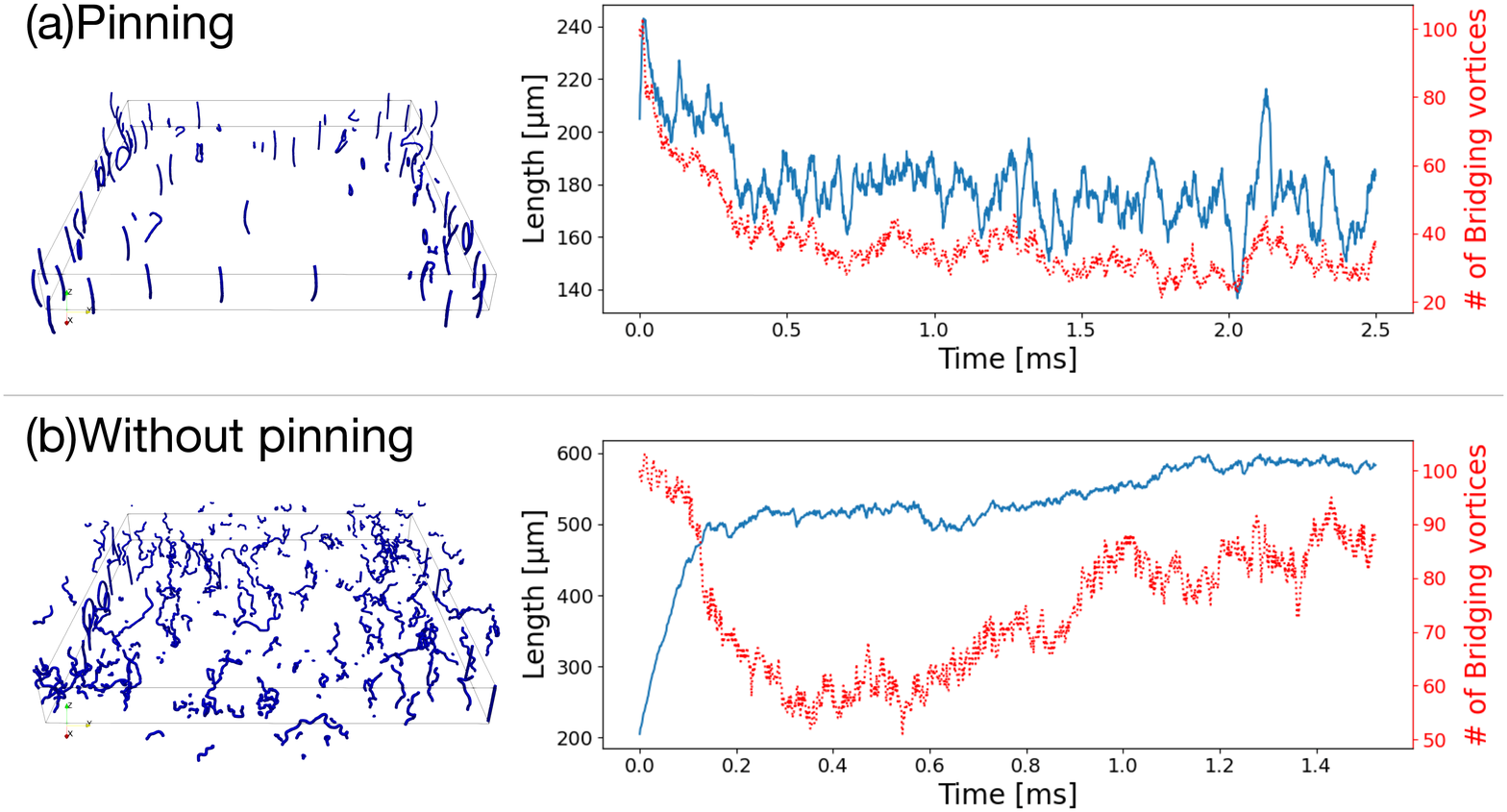}
	\caption{\label{fig:multiplecomp}Time developments of the vortex line length (solid line) and the number of vortices (dotted line) of the multiple vortex case (a) with pinning and (b) without pinning in the critical angle model. }
\end{figure}

\begin{figure}
	\includegraphics[width=\linewidth]{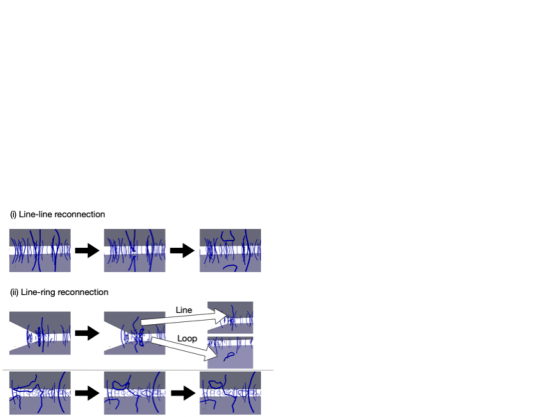}
	\caption{\label{fig:dissipation}Snapshots representing dissipation mechanism in the multiple vortex simulation in the critical angle model. In (i), a pair of  antiparallel vortices reconnect with each other and split into two half loops attached to the boundaries. The half loops eventually dissipate by RSID. The upper of (ii) shows the collision of an injected vortex ring with a pinned line vortex. The reconnection event induces fluctuations and/or kinks on the line and the ring. The fluctuating loop becomes a half loop when in contact with a boundary and  eventually dissipates away by RSID. The kinked structure along the line also dissipates by RSID. The lower of (ii) depicts the pinned line vortex under heavy fluctuations after multiple collisions with vortex rings. This vortex line can emit vortex loops through self-reconnection, which also get dissipated eventually.}
\end{figure}

It is interesting to note that despite the continuous injection of vortex rings, a dense tangle is not generated in the pinned case, as shown in Fig. \ref{fig:multiplecomp}(a) (also, see Ref. \cite{injection_sup}). The total vortex line length in the volume $V$ actually decreases from the initial value of $200\ \mathrm{\mu m}$ to a statistically steady value of $180\ \mathrm{\mu m}$. The number of vortices bridging the boundaries also decreases to about 30 within the volume, inhabiting mostly near the edge of the boundary. Note that the calculations for the total vortex line length and the number of bridging vortices are performed only within the specified volume $V$. 
During the simulation, the vortices frequently undergo reconnections and topological changes. The vortices that undergo reconnections tend to adopt configurations that are more susceptible to dissipation through the RSID mechanism. We identified two types of dissipation mechanisms associated with the reconnections: (i) line-line reconnection, (ii) line-ring reconnection. The examples of these events are shown in Fig.~\ref{fig:dissipation}.
(i) Line-line reconnection: A pair of antiparallel vortices reconnect with each other and split into a pair of half loops that are attached to the boundaries. These loops gradually shrink due to RSID, as they tilt under the influence of their self-induced velocity. Eventually, they completely disappear.
(ii) Line-ring reconnection: An injected vortex ring reconnects with a line vortex. After the reconnection, they form a fluctuating or kinked loop and line \cite{Tsubota2000}. While the kinked structure along the line vortex is often dissipated through RSID, it sometimes experiences significant fluctuations due to the multiple collisions with vortex rings in a short time. In such cases, the highly fluctuating line vortex can emit a vortex loop through self-reconnection. The loops generated by the reconnection of a ring and a line, as well as those generated by the self-reconnection of a line vortex, tend to approach the boundary and eventually attach to it, experiencing the dissipation process similar to the case (i). Some loops escape from the system without interacting with the boundaries and consequently contribute to dissipation.

\begin{figure}
	\includegraphics[width=\linewidth]{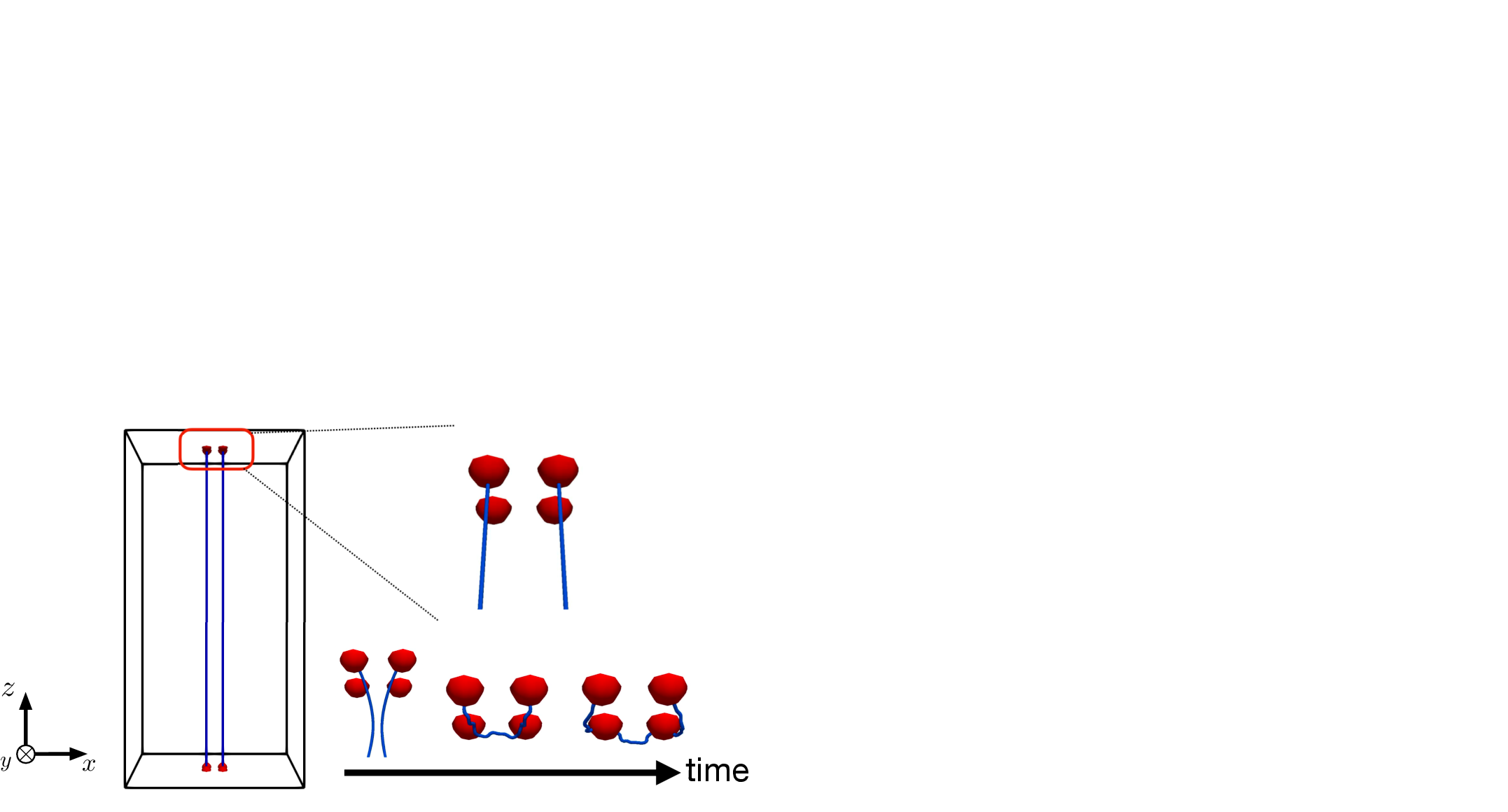}
	\caption{\label{fig:pinninginitial}Vortex dynamics in the hemispherical pinning site model. Four pinning sites are placed on each boundary.  Two antiparallel line vortices whose ends are pinned are initialized. The upper plate is in motion. The rotating line vortices reconnect with each other and split into two half loops attached to each boundary and a closed loop separated in the gap. The half loop subsequently reconnects with the other bumps, splits into smaller vortices and finally dissipates. }
\end{figure}

Let us examine the behavior near the wall in the (i) line-line reconnection and (ii) line-ring reconnection scenarios. To provide a more accurate description, we employ the hemispherical pinning site model, which offers greater detail than the critical angle model. In this model, we place four pinning sites on each boundary and initialize an antiparallel vortex line pair whose ends are pinned, as shown on the left side of Fig.~\ref{fig:pinninginitial}. The upper boundary is in motion.  The snapshots illustrating the process are displayed on the lower right side of Fig.~\ref{fig:pinninginitial}. When the upper boundary starts to oscillate, the termination point of the vortex line moves around on the pinning site while maintaining pinning. The antiparallel vortices are stretched and rotate in the opposite directions, reconnect with each other, and split into two half loops attached to the pinning sites -- one on each boundary -- and a closed loop between the boundaries. The half loop subsequently reconnects with other pinning sites and split into the smaller ones. The splitted vortices propagate on the surface of the boundary and do not affect the vortex dynamics so much. This continuous splitting and reconnection process leads to the dissipation. These results support the concept of RSID described in the critical angle model. We note that this simulation does not perfectly satisfy the boundary condition, as the small distance between the sites is comparable to the radius of the pinning site, thereby breaking the boundary condition at each site. Nevertheless, we believe that this result is qualitatively correct.

In addition, we simulate the multiple vortex case without pinning, where the boundaries are just flat. The dynamics observed in this case is completely different from that with pinning. The snapshots and the time development of the vortex line length in the absence of pinning are illustrated in Fig.~\ref{fig:multiplecomp}(b) (also, see Ref. \cite{injectionnopin_sup}). In the presence of pinning, the vortex line length decreases from the initial value to a statistically steady state (approximately $180\ \mathrm{\mu m}$). However, in the absence of pinning, the vortex line length increases significantly, and a dense tangle is developed (the length is approximately $600\ \mathrm{\mu m}$). Obviously, the number of vortices is larger compared to the case with pinning. Thus, pinning can effectively prevent the generation of turbulence.  This is consistent with the experimental observation \cite{Barquist2022} that they could not induce turbulence with the MEMS oscillator while the turbulence could be readily generated with the QTF.

\subsection{Comparison with experiment}\label{sec:comp}
\subsubsection{Damping mechanism}
Our simulations provide overall consistent pictures with the observations and interpretations made by Barquist {\it et al.}~\cite{Barquist2020,Barquist2022}.   At the heart is the resonant energy transfer from the oscillator to the pinned vortex line exciting the Kelvin waves.  In their experiment, the oscillation frequency was accidentally matched with the first mode of the Kelvin wave for the given geometry.  We confirm that the coupling between the two mechanical systems (oscillator + vortices) is throttled when the two frequencies are not matched.  Furthermore, by introducing pinning-depinning effects in our models we establish the proper dissipation in the vortex system, which appears as damping in the oscillator (oscillating boundary).  Therefore, in the presence of pinning {\it i.e.} dissipation, the oscillator and the pinned vortex system reach a steady state, preventing vortex tangle.  Our simulation also successfully showed how the injected rings interact with the trapped vortices and the boundaries. 

In an attempt to establish a connection between the damping force measured in the experiment and the numerical results, we adopt the assumption that the damping force on the oscillator is caused by the tension of the pinned vortex lines \cite{Barquist2020}.  This allows us to estimate the damping force, which is directly proportional to the oscillation velocity. 
The tension is equivalent to a vortex energy per unit length \cite{Donnelly,Adams1985}
\begin{equation}\label{eq:tension}
	\epsilon=\frac{\rho_\mathrm{s}\kappa^2}{4\pi}\ln(R_0/a_0),
\end{equation}
where $\rho_\mathrm{s}$ and $R_0$ are the superfluid density and a characteristic length. We choose $L=2\ \mathrm{\mu m}$ between the boundaries as $R_0$, which produces $\epsilon = 1.12\ \mathrm{pN}$.  We calculate the force $F_{x}$ by multiplying the $x$ component of the unit tangent vector on the edge of the vortex attached to the pinning site on the upper boundary (in the hemispherical pinning site model) or the upper boundary itself (in the critical angle model). It is important to note that we calculate the force for both the real and image vortex, taking into account the boundary conditions in our analysis.

We can compare the time development of the force from a single vortex in the two models. Let us start with the critical angle model. As shown in Fig.~\ref{fig:force_comp}(a), once the boundary is set in motion, the vortex is gradually stretched and reaches a steady state as previously described. In this process, the angle between the vortex and the boundary decreases to the critical angle, $\theta_{c}$. Consequently, the amplitude of the force approaches a constant value $F_{x} \approx 2\epsilon\cos{\theta_{c}}$. The oscillation in the force is due to the vortex rotation around its symmetry axis. On the other hand, in the hemispherical pinning model as shown  Fig.~\ref{fig:force_comp} (b), the force gradually increases but when the vortex is sufficiently stretched, it becomes depinned and the force abruptly disappears. Note that  Fig.~\ref{fig:force_comp}(b) shows the results in a much narrower time duration. 

In both models, the force strongly depends on the oscillation frequency as expected. When the oscillation frequency is close to the resonant frequency, the vortex experiences significant stretching; the angle becomes smaller, leading to a higher force. (see Fig.~\ref{fig:force_comp}) . However, when the frequency deviates from any resonant frequencies, the vortex is not stretched as much, resulting in a relatively smaller force. It is worth to note that even in these non-resonant frequencies, the force oscillation can still excite a mode at other than the actual oscillation frequency.   For example, the spectra for $10000\ \mathrm{Hz}$ and $40000\ \mathrm{Hz}$ exhibit the peaks at their own oscillation frequencies as well as at the Kelvin wave mode frequencies as shown in Fig.~\ref{fig:angle_force_fourier}. In contrast, the spectrum for $23625\ \mathrm{Hz}$ shows only a high resonant peak.

The force shown in the simulation is composed of two quadrature components: $F_{x} = F_1\sin(2\pi f t) + F_2\cos(2\pi f t)$.  In our simulation, $F_{1}$ and $F_{2}$ correspond to the component proportional to the acceleration and the velocity, respectively. Therefore, $F_{2}$ is the damping force. Based on this we obtained the values for $F_1=-0.63\ \mathrm{pN}$ and $F_2=-1.44\ \mathrm{pN}$ from the 23625~Hz force oscillation in the critical angle model. Using the velocity amplitude $v_{0} = 2\pi f A$, the damping coefficient is calculated to be $\eta = 9.7\times10^{-11}\ \mathrm{kg/s}$ where $F_{x} = \eta v_{0}$. This value is an order of magnitude larger than the rough estimation of $5.4\times10^{-13}\ \mathrm{kg/s}$, given in Ref. \cite{Barquist2022}: 


\begin{figure}
	\includegraphics[width=\linewidth]{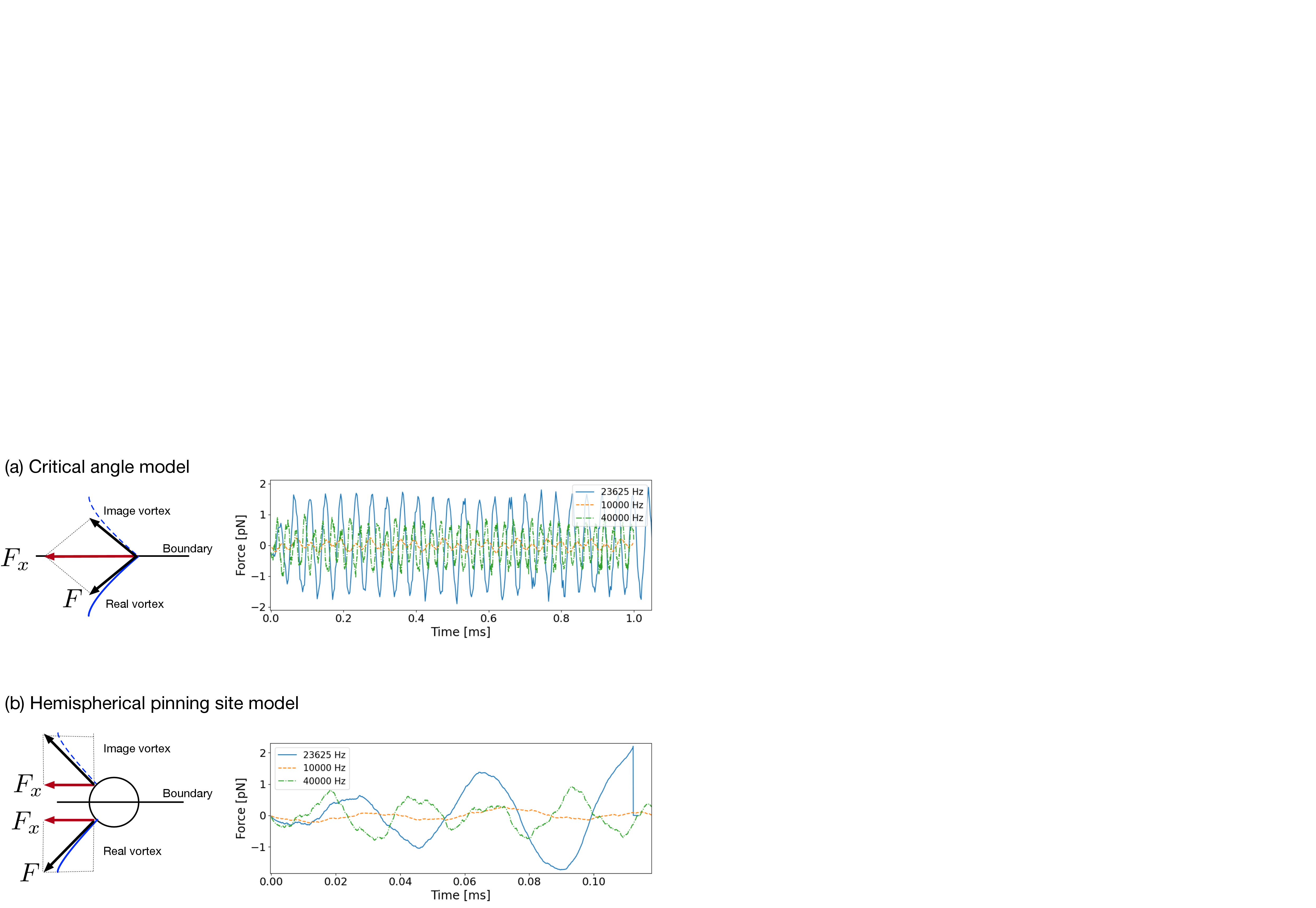}
	\caption{\label{fig:force_comp}Force acting on the boundary along the direction of the oscillation in the single vortex case in (a) the critical angle model (Fig. \ref{fig:singlesteady}) and (b) the hemispherical pinning site model (Fig. \ref{fig:singlesite}). The left sides of (a) and (b) are the force diagrams of the tension acting on the boundary by the real and imaginary vortex. The right sides of (a) and (b) show the time development of the force.}
\end{figure}

\begin{figure}
	\includegraphics[width=\linewidth]{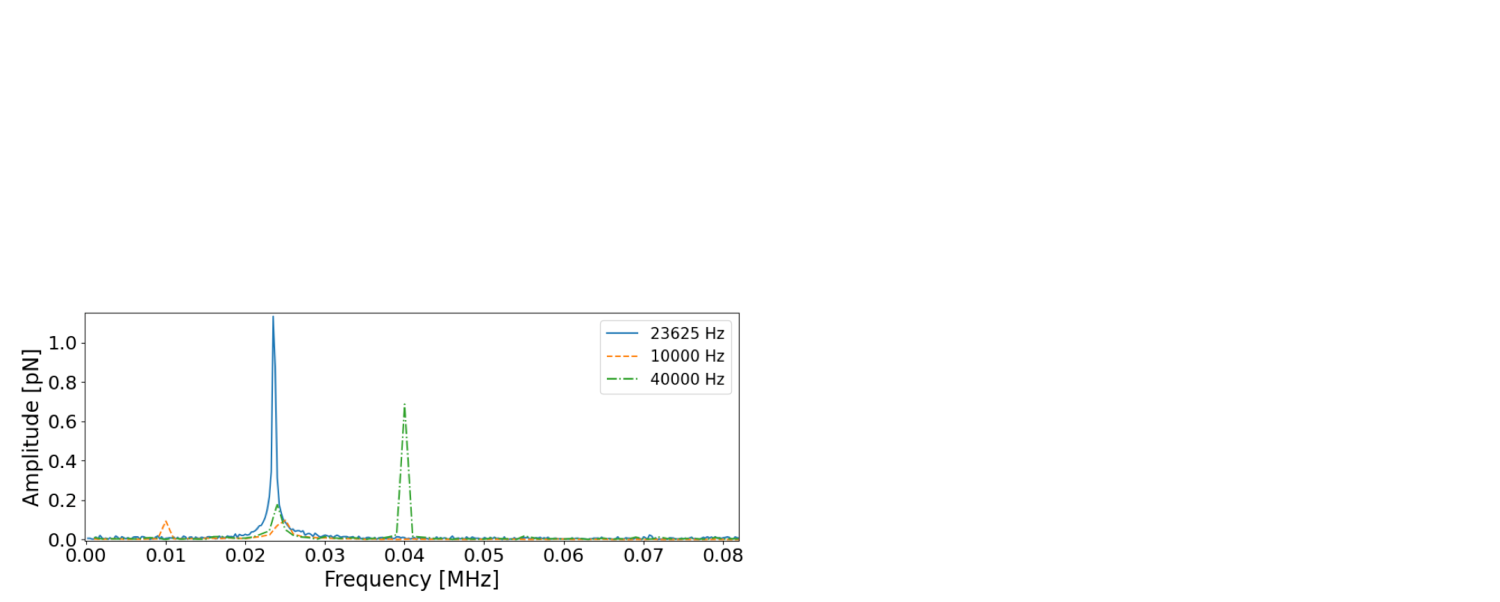}
	\caption{\label{fig:angle_force_fourier}Fourier transform of the time development of the force in Fig. \ref{fig:force_comp} (a).} 
\end{figure}

\begin{equation}\label{eq:coefficient}
	\eta=2\frac{\epsilon f}{v_0^2}\delta l.
\end{equation}
where $\delta l$ is the vortex length stretching in a cycle of oscillation. 

\begin{figure}
	\includegraphics[width=\linewidth]{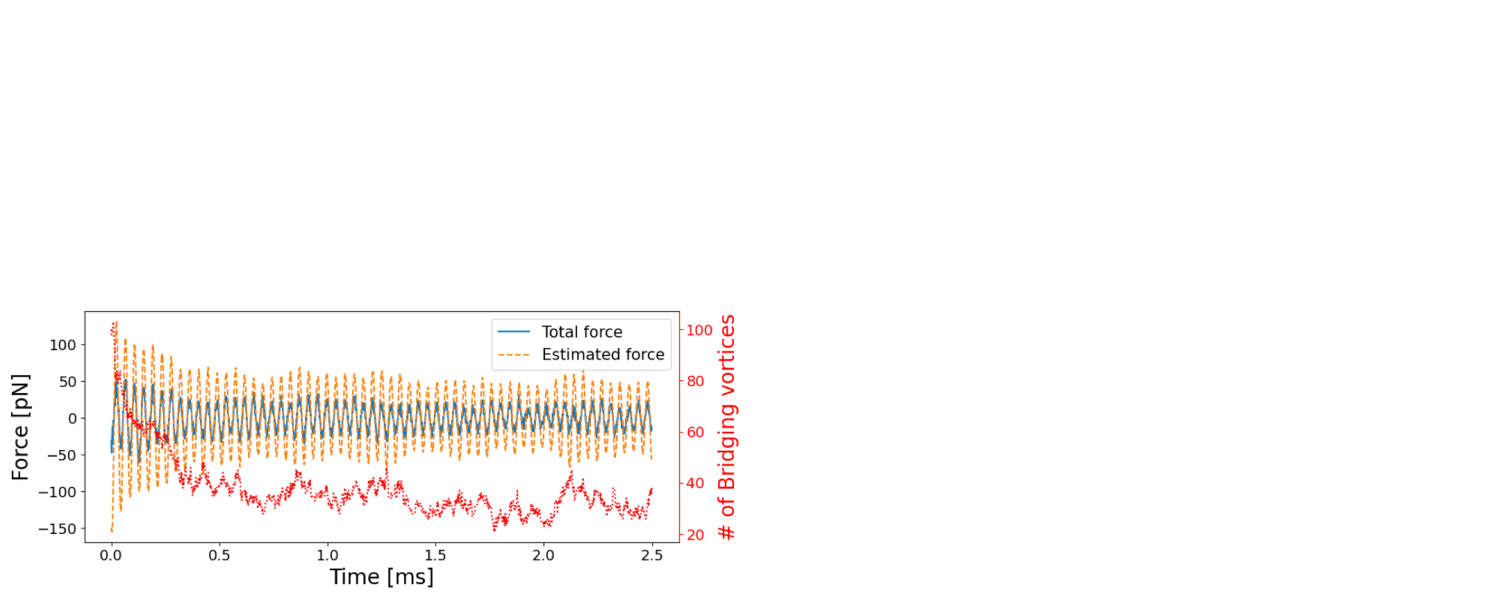}
	\caption{\label{fig:injected_force}Time dependence of the force acting on the plate, $F_{x}$, and the number of the bridging vortices (dotted line) in the critical angle model. The estimated force is calculated by multiplying the force in the single vortex case with the oscillation frequency $23625\ \mathrm{Hz}$ and the number of the bridging vortices shown by a dotted line.}
\end{figure}

\begin{figure}
	\includegraphics[width=\linewidth]{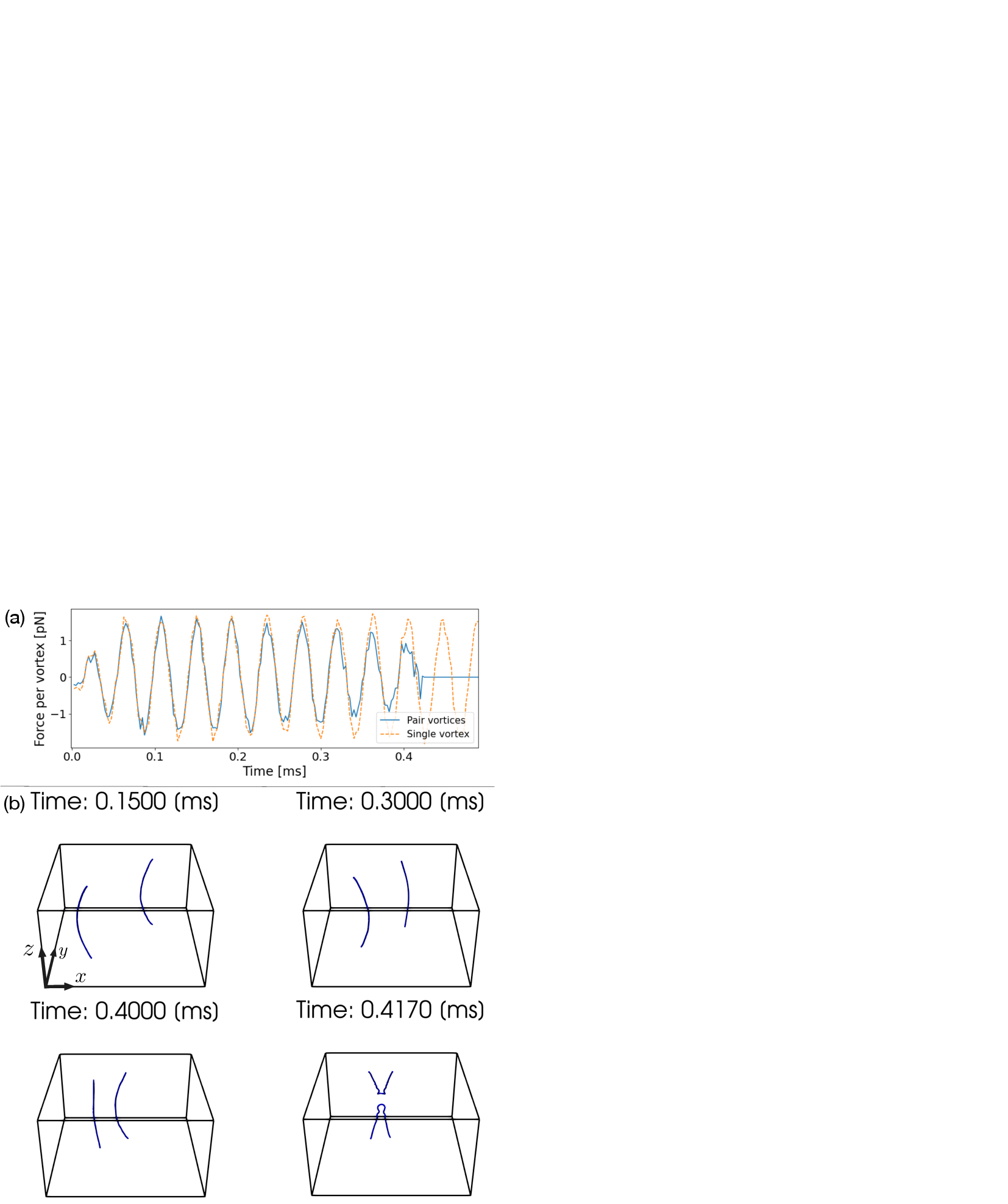}
	\caption{\label{fig:pair_vortex}(a)Comparison of the force in the pair vortex case and the single vortex case (Fig. \ref{fig:force_comp} (a)) by the critical angle model.  The lower figures shows the dynamics of the pair vortices with the oscillation frequency $23625\ \mathrm{Hz}$. In the pair vortex case, the shown force is calculated by dividing the total force by two. (b) Time development of the pair vortices. The virtual rectangle depicted by the black lines is $3.6\ \mathrm{\mu m}$ in $x$ and $y$ directions and $2\ \mathrm{\mu m}$ in $z$ direction, and its center is the origin. They initially rotates independently by their self-induced velocity. They gradually get closer in interaction, and the interaction becomes more dominant compared to rotation by self-induced velocity. Finally, the vortices reconnect with each other, and dissipated by RSID.}
\end{figure}

In the case of multiple vortices, Fig.~\ref{fig:injected_force} shows the time dependence of the force and the number of the bridging vortices. The orange dashed line represents the estimated total force obtained from the single vortex simulation by mutiplying the total number of vortices. The estimated force turns out to be greater than the total force observed in the multiple vortex simulation (blue solid line). It is notable that strong fluctuations in the the number of the bridging vortices are present, which is directly related to the processes described in Sec.~{\ref{sec:multi}}.  The number of bridging vortices fluctuates due to the population of the new vortices from the injected rings and the annihilation of the existing vortices.  The enhanced noise in the presence of vortex ring injection was observed in the experiment in the form of the phase noise \cite{Barquist2022}.  The difference between the estimated force and the total force in the multiple vortex case can be attributed to vortex interaction.  To illustrate this, we consider an example involving an antiparallel rectilinear vortex pair. Each vortex is located at coordinates $(x,y)=(0.9\ \mathrm{\mu m},0.9\ \mathrm{\mu m})$ and $(-0.9\ \mathrm{\mu m},-0.9\ \mathrm{\mu m})$, respectively. These vortices are pinned between the boundaries, and the upper boundary is oscillated. Figure \ref{fig:pair_vortex}(a) illustrates the force per vortex in the pair vortex case compared to the force in the single vortex case.  The time development of the dynamics is shown in Fig.\ref{fig:pair_vortex}(b). Initially, each vortex moves independently because their interaction is weak, and the force acting on the plate is approximately equal to their sum. However, as the amplitude of the Kelvin wave increases and the distance between the vortex pair decreases, the interaction between the two vortices gets dominant, causing disturbances in their rotational motion.  The phases of the vortex rotation became out of sync, and the orientations of the forces by the vortices are not aligned anymore. As the result, the total force exerted by the vortex pair decreases. This local mechanism occurs extensively over the whole vortices, leading to the discrepancy between the total force and the estimated force.

In this study, we have exclusively concentrated our attention on vortices residing within the gap between the plate and the substrate. Nevertheless, within the real system, it is plausible that vortices pinned to the sides facing the bulk could potentially impact the motion of the plate. We consider that their influence is negligible. Typically, these vortices have much longer than those within the gap, and they are energetically less favorable in a vortex nucleaction process. Furthermore, the oscillation frequnecy generally does not match with their Kelvin wave resonance modes. As illustrated in Figure \ref{fig:force_comp}, the tension with off-resonance oscillation has a relatively minor effect. Consequently, these vortices exert only a modest influence on the damping.

\subsubsection{Origin of the force hysteresis}
\begin{figure}
	\includegraphics[width=\linewidth]{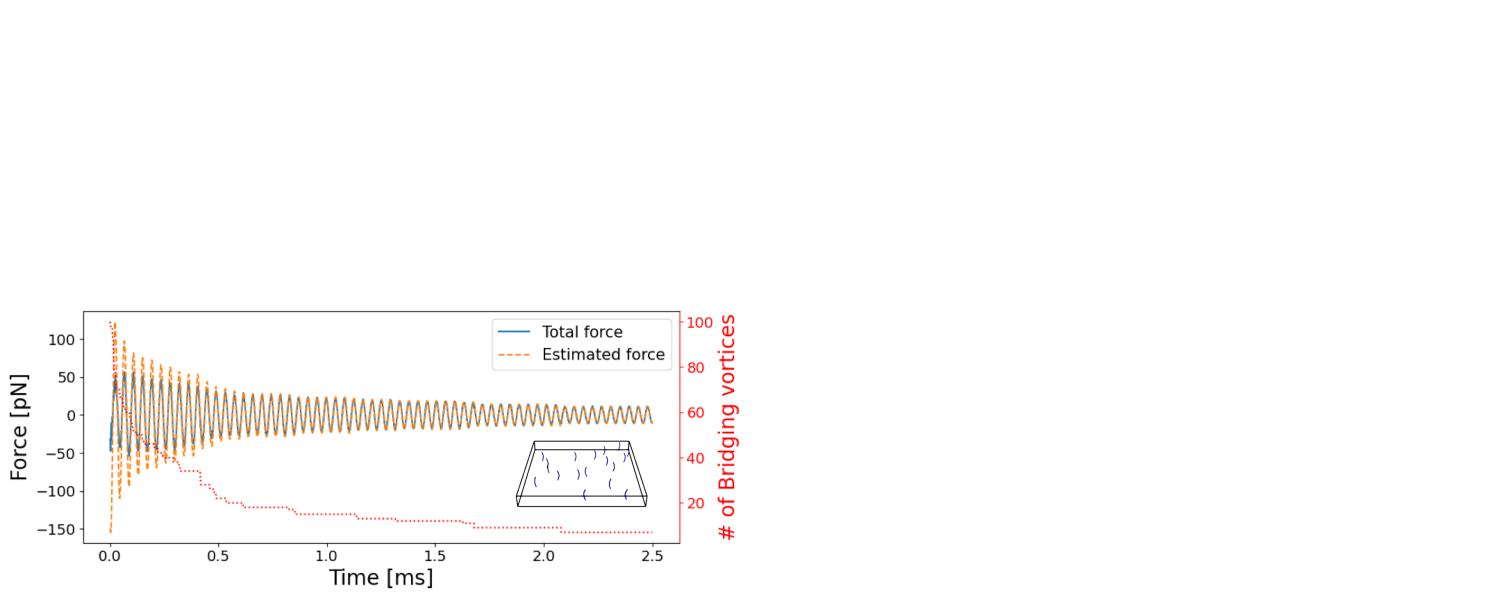}
	\caption{\label{fig:no_injected_force} Total force acting on the plate along the $x$ direction, estimated force and number of bridging vortices (dotted line) in the multiple vortex simulation without injection by the critical angle model.  The inset of this figure is a snapshot of the dynamics at the time $0.6455\ \mathrm{ms}$, where the vortices are degenerate.} 
\end{figure}

We propose that the force hysteresis observed in the experiment can be explained by the multiple vortex simulation (Sec. {\ref{sec:multi}}). The force hysteresis is observed only when vortices are not externally injected. The Florida group suggested that this implies that the damping is directly related to the number of pinned vortices, and the number of vortices decreases for some reason when the oscillator was driven hard in the absence of external vortex injection.  We performed a multiple vortex simulation without injection. The total force and the estimated force are presented in Fig.~\ref{fig:no_injected_force}.  Unlike the case with the vortex injection, the amplitude of the total force is smaller than that of the estimated force in the beginning, but as time goes by, the difference between them disappears. This is due to the decrease in the vortex density, which allows each vortex to move independently without reconnection processes. This process leads to a gradual decrease in the total force and the number of vortices. Such a mechanism contributes to the observed force hysteresis.  Furthermore, the fluctuations in the force and the number of bridging vortices are significantly smaller compared to the case with the vortex injection.


\section{Conclusion}\label{sec:conclusion}
We conducted extensive numerical simulations to study the dynamics of quantized vortex systems with pinning within the framework of two pinning models: the critical angle model and the hemispherical pinning site model. The critical angle model, although phenomenological, is suitable for describing the vortex dynamics with rough solid boundaries. Dissipation mechanism is successfully incorporated in this model even at 0 K through reconnection of vortices with rough boundaries, referred to as rough surface-induced dissipation (RSID).  On the other hand, the hemispherical pinning site model provides a more accurate representation of pinning-depinning process by correctly solving the boundary conditions. Although his model does not inherently include vortex dissipation, it can lead to dissipation when vortices interact with rough boundaries, as illustrated in Fig. \ref{fig:pinninginitial}. \\

Using these models, we performed calculations to simulate vortex dynamics with pinning in the geometry closely emulating the experimental set-up.  We confirmed that the dissipation of the pinned vortices is the main damping mechanism of the oscillator at zero temperature when the oscillator was driven at a frequency matched to a Kelvin wave mode of the vortices.  In the presence of multiple vortices, the pinned vortices interacted with each other and also with the injected vortex rings, and experienced dissipation due to a combination of interaction and RSID. This dissipation prevented the development of turbulence in the system, which would otherwise occur in the absence of pinning. Additionally, we performed a multiple vortex simulation without any injection. In this case, the number of vortices and the force on the plate decreased over time, which may explain the origin of the force hysteresis phenomenon.

\begin{acknowledgments}
This work is supported by JST SPRING, Grant No.~JPMJSP2139 and JSPS KAKENHI Grant No.~JP23KJ1832 (TN); JSPS KAKENHI Grant No.~JP23K03305 (MT); NSF Grant No.~DMR-1708818 (YL). The authors acknowledge University of Florida Research Computing for providing computational resources and support that have contributed to the research results reported in this publication. 
\end{acknowledgments}

\bibliography{ref.bib}

\end{document}



\title{Dynamics of pinned quantized vortices in superfluid $^4$He in a microelectromechanical oscillator}


\author{Tomo Nakagawa}
\affiliation{Department of Physics, Osaka City University, 3-3-138 Sugimoto, 558-8585 Osaka, Japan}
\author{Makoto Tsubota}
\affiliation{Department of Physics, Nambu Yoichiro Institute of Theoretical and Experimental Physics(NITEP), Osaka Metropolitan University, 3-3-138 Sugimoto, 558-8585 Osaka, Japan}
\author{Keegan Gunther}
\affiliation{Department of Physics, University of Florida, Gainesville, FL 32611, USA}
\author{Yoonseok Lee}
\affiliation{Department of Physics, University of Florida, Gainesville, FL 32611, USA}





\maketitle

\section{Dependence of the critical angle in the critical angle model}\label{sec:introduction}
In the critical angle model, the vortex dynamics depends on the chosen critical angle. When the vortex is strongly pinned to the wall bump, it tilts with oscillation and moves closer to the wall. However, if the vortex gets too close to the wall without interacting or reconnecting with a sufficiently rough surface, this state is undesirable, because this contradicts the reconnection criterion. Some cutoff within which vortices should be reconnected needs to be introduced. In our study, we selected a critical angle of $\pi/6$, which is worse to take too large or too small value, to account for this behavior.

\begin{figure}
	\includegraphics[width=0.78\linewidth]{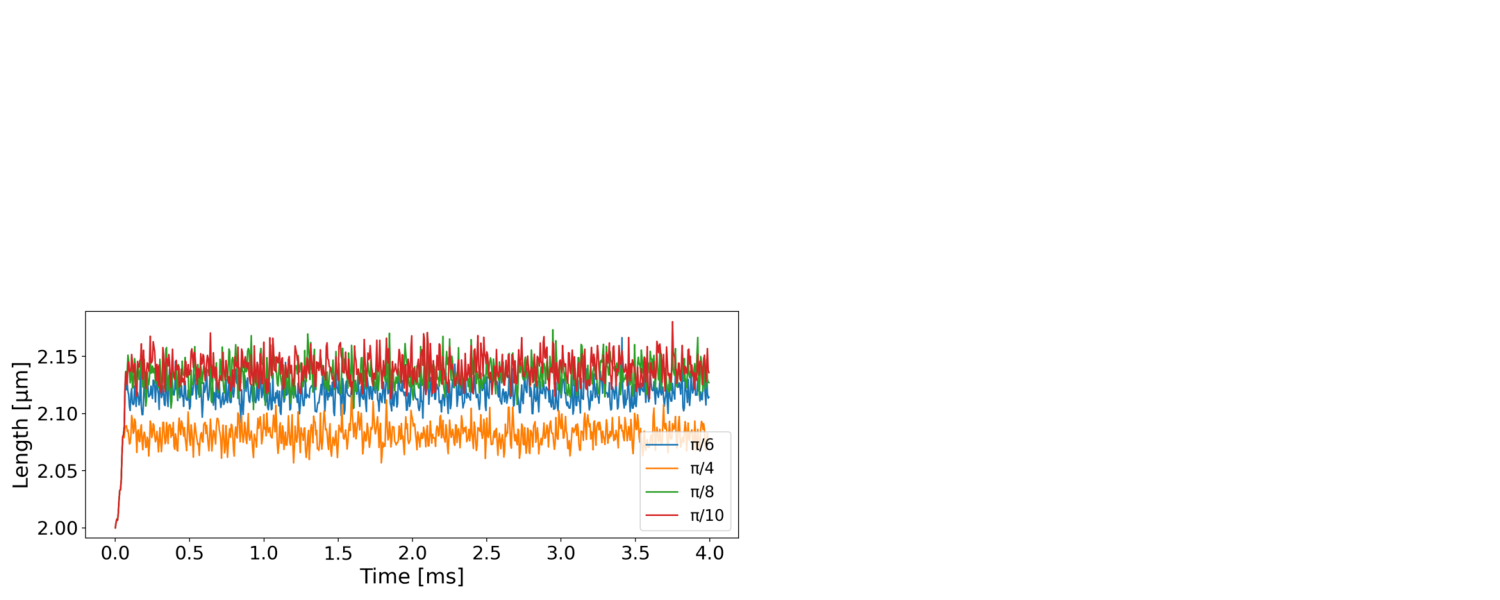}
	\caption{\label{fig:angle_dependence}Critical angle dependence of vortex line length in the single vortex case. }
\end{figure}

We conducted calculations for different critical angles ranging from $\pi/10$ to $\pi/4$ in the case of a single vortex. Although the lengths of the vortices in the steady state slightly differ, all vortices exhibit similar qualitative behavior. The vortex line length with the angle of $\pi/6$, $\pi/8$ and $\pi/10$ seems to converge, although the dynamics with the angle of $\pi/4$ is clearly different.  Then, we use $\pi/6$ as the critical angle.

\section{Pinning site size dependence of vortex dynamics in the spherical pinning site model }

\begin{figure}
	\includegraphics[width=0.78\linewidth]{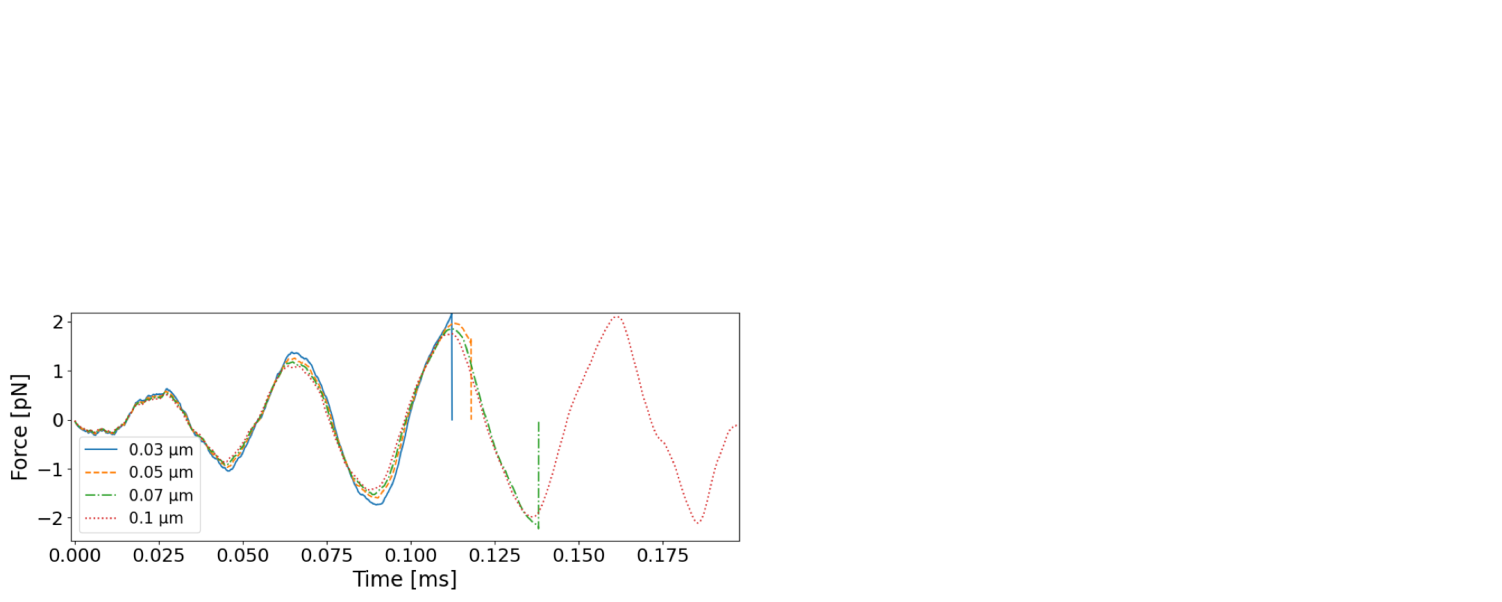}
	\caption{\label{fig:angle_size}The time development of the force in the hemispherical pinning site model. The legend indicates the pinning site size $a$. Plottings are terminated when the pinned vortex is depinned.}
\end{figure}

In the main manuscript, our discussion of the dynamics is limited in the spherical pinning site model with a pinning site of $a=0.03\ \mathrm{\mu m}$.
 Figure \ref{fig:angle_size} shows the time development of the force acting on the site in the direction of oscillation with four different values of $a$. Note that the oscillation frequency $f$ and amplitude $A$ are held constant at $23625\ \mathrm{Hz}$ and $0.1\ \mathrm{\mu m}$, respectively.  The moment when the force abruptly diminishes to zero signifies the onset of depinning. 

We observe that as the pinning site radius increases, the vortex becomes less likely to be depinned. The phenomenon of vortex depinning in the hemispherical pinning site model occurs due to a reconnection event between the vortex and a flat solid boundary. This reconnection event happens when the distance between the vortex point on the site (strictly, its adjacent point) and the solid boundary falls below the spatial resolution $\Delta\xi_\mathrm{min}$. Consequently, the effective critical angle for vortex depinning($\sim\arcsin(a/\Delta\xi_\mathrm{min})$), decreases as the pinning site radius $a$ increases.


Furthermore, it is worth noting that the vortex pinned on a site with a radius of $a=0.1\ \mathrm{\mu m}$ remained securely pinned throughout the computational period we conducted, which is extended for approximately $0.35\ \mathrm{ms}$. 
This is probably because the pinning site radius $a$ becomes comparable with its amplitude $A$.
As a result, the vortex point on the site did not approach a flat boundary during the simulation, leading to its stable pinning.
\bibliography{ref.bib}